\documentclass[aip,amsmath,amssymb,reprint,citeautoscript,nofootinbib]{revtex4-1}

\usepackage{pgfplots}
\usepackage{graphicx}
\usepackage{dcolumn}
\usepackage{bm}
\usepackage{float}
\usepackage{tikz}
\usepackage{environ}
\makeatletter
\newsavebox{\measure@tikzpicture}
\NewEnviron{scaletikzpicturetowidth}[1]{%
  \def\tikz@width{#1}%
  \begin{lrbox}{\measure@tikzpicture}%
  \BODY
  \end{lrbox}%
  \pgfmathparse{#1/\wd\measure@tikzpicture}%
  \BODY
}
\makeatother
\usetikzlibrary{calc,angles,positioning,intersections}
\usetikzlibrary{decorations.pathreplacing,quotes}
\usetikzlibrary{calc,angles,positioning,intersections}
\usepackage{graphicx}
\usepackage{dcolumn}
\usepackage{bm}
\usepackage[normalem]{ulem} 

\graphicspath{{fig/}}
\usepackage[utf8]{inputenc}
\usepackage[T1]{fontenc}
\usepackage{mathptmx}
\pgfplotsset{compat=newest}

\newcommand{\xinit}{x_{\text{initial}}}
\newcommand{\xfinal}{x_{\text{final}}}
\newcommand{\acrit}{a_{\text{crit}}}
\newcommand{\atricrit}{a_{\text{tricrit}}}

\begin{document}

\preprint{AIP/123-QED}

\title{A coupled oscillator model for the origin of bimodality and multimodality}

\author{J.D. Johnson}
\author{D.M. Abrams}%
\affiliation{ 
Department of Engineering Sciences and Applied Mathematics, Northwestern University \\ McCormick School of Engineering and Applied Science \\
2145 Sheridan Road
Evanston, IL 60208
}%

\date{\today}

\begin{abstract}
Perhaps because of the elegance of the central limit theorem, it is often assumed that distributions in nature will approach singly-peaked, unimodal shapes reminiscent of the Gaussian normal distribution.  However, many systems behave differently, with variables following apparently bimodal or multimodal distributions. Here we argue that multimodality may emerge naturally as a result of repulsive or inhibitory coupling dynamics, and we show rigorously how it emerges for a broad class of coupling functions in variants of the paradigmatic Kuramoto model.
\end{abstract}

\maketitle

\begin{quotation}
In this paper we employ oscillators as a test system for understanding how bimodality---the splitting of oscillators into two rather than one cluster---may emerge as a result of coupling between interacting units.  We present numerical and analytical results showing that repulsive coupling can lead to bimodality (or multimodality) for a wide range of detailed interaction dynamics. 
\end{quotation}

\section{\label{sec:level1}Introduction} 

Synchronization is a widespread phenomenon observed in biological \cite{saigusa2008amoebae,myung2015gaba,taylor2009dynamical}, chemical \cite{toiya2010synchronization,vaidyanathan2015anti,li2003study}, physical \cite{pantaleone2002synchronization,yoshida2000phase,henk2003synchronization,ulrichs2009synchronization}, and social settings \cite{repp2005sensorimotor,kirschner2009joint,de2009inference,pluchino2006compromise}.
A paradigmatic mathematical model that can explain synchronization in many contexts is the Kuramoto model \cite{kuramoto2003chemical,kuramoto1975self,strogatz2000kuramoto,acebron2005kuramoto,moreno2004synchronization}. Much work has been done on understanding the complex and surprising dynamics of the Kuramoto model and its variants, but the vast majority of that research focuses on the case of attractive coupling; here we are interested in the case where the coupling is repulsive.


Repulsive (or inhibitory) coupling is of physical interest as it arises frequently in the context of neuronal networks (e.g., see refs.~\onlinecite{myung2015gaba, wang2011synchronous}), chemical interactions (e.g., refs.~\onlinecite{toiya2010synchronization,epstein1991nonlinear,hohmann1998learning}), and many other systems (see refs.~\onlinecite{koseska2007inherent,ullner2007multistability,ullner2008multistability,marvel2009energy,laje2002diversity,balazsi2001synchronization}).  Some coupled oscillator models have examined repulsive coupling: Giver et al.~developed a local variant of the Kuramoto model with repulsive coupling based on the interaction between water micro-droplets with reactants of the Belousov-Zhabotinsky reaction~\cite{giver2011phase}.  Hong and Strogatz developed two variants of the Kuramoto model that involved mixes of positive and negative coupling~\cite{hong2011kuramoto,hong2012mean}. 


\begin{figure}[t!]
    \centering
    \includegraphics[width=0.99\columnwidth]{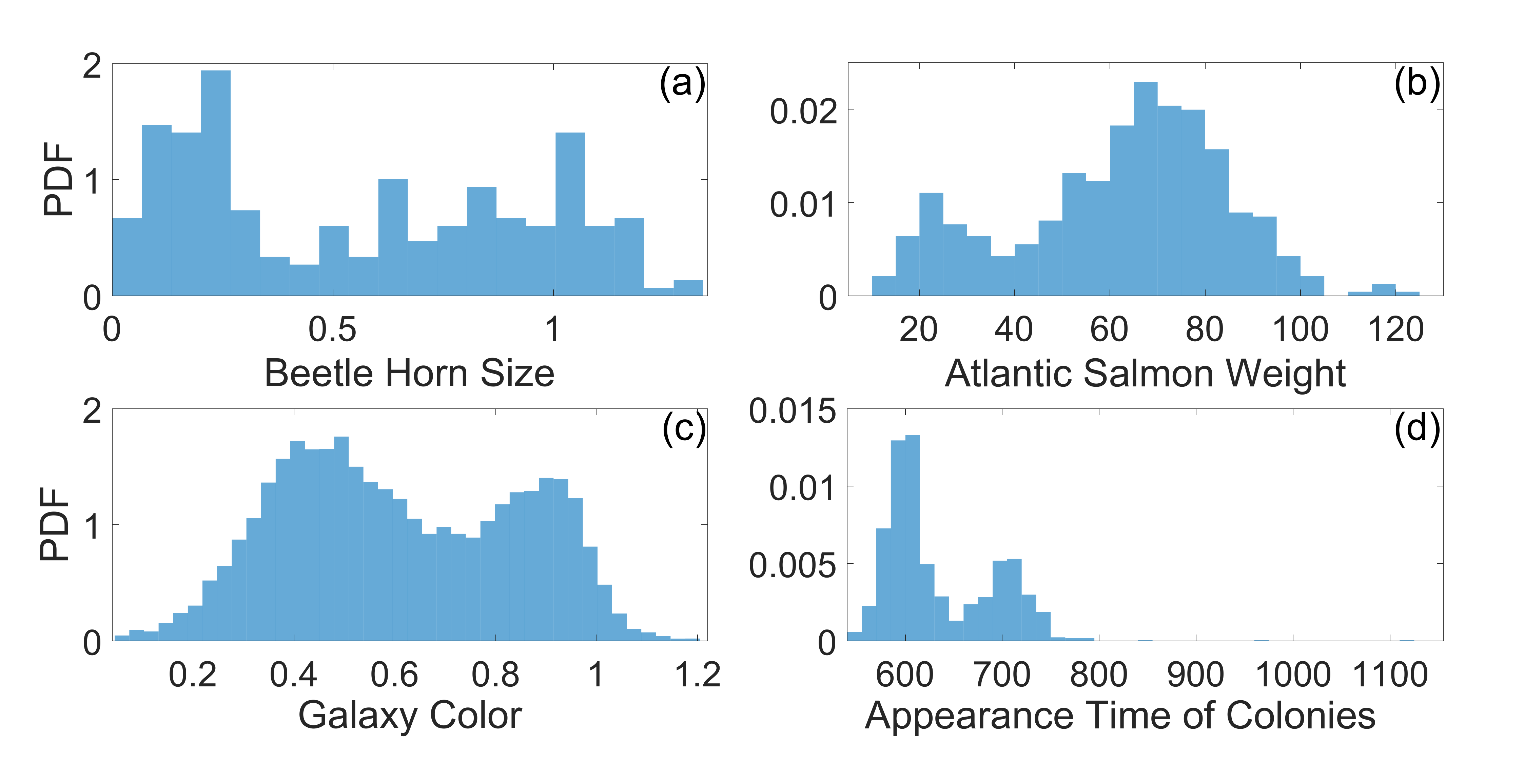}
    \caption{\textbf{Selected examples of bimodality.} Histograms (normalized) for (a) size of beetle horns [mm], \cite{emlen1996,clifton2016,clifton2016data} (b) Atlantic salmon body mass [g]\cite{damsgaard2019proactive,dryad_dat} (c) color of galaxies at redshift 0.1\cite{blanton2003broadband,abazajian2003first,baldry2004} (d) inverse growth rates of bacteria [$\textrm{min}^{-1}$] \cite{ronin2017long,goldberg2014systematic}. }
    \label{fig:BimodExamples}
\end{figure}

The relationship between network structure and repulsive coupling has also been analyzed, with Levnaji\'c\cite{levnajic2011emergent, levnajic2012evolutionary} showing that, given the network coupling structure, many different phase configurations can arise. 
Recently, it has been shown that synchronization can arise in both repulsive and attractive coupling scenarios subject to common noise \cite{pimenova2016interplay,nagai2010noise,gil2010common,gong2019repulsively}. Gong et al.\cite{gong2019repulsively}, inspired by the work of Gil et al.\cite{gil2010common}, studied instances where common noise can lead to clustering in the phase distribution of oscillators for repulsive coupling. 

Nakamura et al.\cite{nakamura1994clustering} investigated the effect of time-delayed nearest-neighbor coupling in the Kuramoto model and found that it could lead to the development of clustered states for both attractive and repulsive coupling. Mishra et al.\cite{mishra2015chimeralike} demonstrated that ``chimeralike'' states could arise with globally coupled Li\'enard systems incorporating both attractive and repulsive mean-field feedback. Yeldesbay et al.\cite{yeldesbay2014chimeralike} established that chimeralike states can arise in the Kuramoto-Sakaguchi model. They also considered a model with oscillators that could be synchronous (attractive coupling) or asynchronous (repulsive coupling) depending on their natural frequencies. They found that in this case a chimera state arises.

Golomb et al.\cite{golomb1992clustering} showed that clustering is possible in a coupled oscillator model with repulsive coupling that is suited for strong interactions between the limit-cycle oscillators. They further provided theory for when a frequency locked stationary phase distribution and when a nonperiodic attractor can arise.


Tsimring et al.\cite{tsimring2005repulsive} showed that heterogeneous globally coupled oscillators obeying the standard Kuramoto model can cluster with all configurations having a zero order parameter, but this clustering breaks down as the number of oscillators increases. They also showed that, with local coupling,  clustering can occur for nonidentical oscillators given sufficiently large coupling strength. 

Closest to the work we present here, Okuda\cite{okuda1993variety} looked at the effect that an arbitrary coupling function may have on oscillators and developed theory as to when an $n$-cluster state, with all clusters being the same size, can arise. He found that harmonics in the coupling function are necessary for clusters to arise.


The central limit theorem \cite{billingsley1995probability} may influence us to expect that distributions in nature should tend to a singly-peaked, unimodal shape akin to the Gaussian normal distribution. Yet, bimodality and multimodality can be observed in biological \cite{chaudhary2016bimodality,seeman1984bimodal,lewis2003dynamics}, social \cite{wu2010evidence,dimaggio1996have,flay1978catastrophe}, and chemical \cite{bonifacio2015topos,reiter1987bimodality,ren2011impact,sun1977chemical} contexts and beyond\cite{dekel2006galaxy,zhang2003bimodality,kuhl1982bimodal} (see Fig.~\ref{fig:BimodExamples} for selected examples). In this paper we demonstrate that multimodality may arise as a result of repulsive or inhibitory coupling dynamics and we give an in-depth explanation of how it can arise for a range of coupling functions.

\section{Model with antisymmetric repulsive coupling} \label{sec:modelintro}

We begin by considering a system of $N$ phase oscillators characterized by natural frequencies $\omega_i,~i=1\ldots N$.  The oscillators are globally coupled with coupling strength $K$ through an interaction function $f$ that depends only on the phase difference between each pair of oscillators:
\begin{align}
    \dot{\theta}_i= \omega_i + \dfrac{K}{N} \sum_{i=1}^N f(\theta_j-\theta_i),\quad i=1,\ldots,N\;.
\end{align}
Here $K>0$ represents attractive coupling and $K<0$ represents repulsive coupling. 

We consider interaction functions $f(u)$, $u \in (-\pi, \pi]$, that satisfy the following conditions:
\begin{subequations} \label{allconditions}
\begin{align} 
    f(0)   &= 0  \label{ZeroZeroCond} \\ 
    f'(0)  &> 0 \label{SlopePos}\\
    f(u)   &=-f(-u) \label{Oddfun} \\
    f'(u)   &\text{ continuous} \label{DiffFun} \\
    f(\pi) &= \lim_{u \to -\pi^+} f(u). \label{PeriodicFun}  
\end{align}
\end{subequations}
These conditions impose: (\ref{ZeroZeroCond}) no coupling effects between oscillators in sync; (\ref{SlopePos}) locally attractive (repulsive) coupling near sync state for $K>0$ ($K<0$); (\ref{Oddfun}) odd interaction function; (\ref{DiffFun}) no discontinuities in $f'(u)$; (\ref{PeriodicFun}) $2\pi$-periodic interaction function on $(-\pi, \pi]$ domain.
We point out that conditions $(\ref{Oddfun})$ and $(\ref{PeriodicFun})$ lead to $f(\pi) = \lim_{u \to -\pi^+} f(u) = 0$.

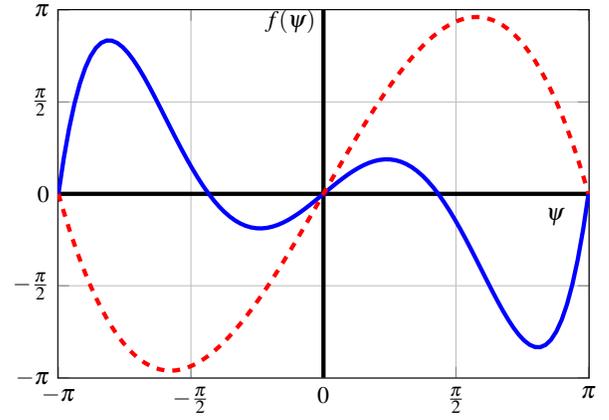
\begin{figure}[t!]
    \centering 
    \begin{tikzpicture}
    \begin{axis}[width = \columnwidth, height = 0.75\columnwidth,grid=both,xmin =-4.01,ymin=-4.01, xmax =4.01, ymax =4.01, xtick={-4,-2,0,2,4}, xticklabels = {$-\pi$, $-\frac{\pi}{2}$,$0$, $\frac{\pi}{2}$,$\pi$},ytick={-4,-2,0,2,4}, yticklabels = {$-\pi$, $-\frac{\pi}{2}$,$0$, $\frac{\pi}{2}$,$\pi$}]
    \node at (-0.5,3.75)  {$f(\psi)$};
    \node at (3.50,-0.5) {$\psi$};
    \draw[ultra thick] (-4,0) -- (4,0);
    \draw[ultra thick] (0,-4) -- (0,4);
  \addplot[samples=100,blue,ultra thick, domain =-4:4]{(\x+4)*\x*(\x-4)*(\x*\x-3)/40)};
    \addplot[samples=100,red,ultra thick, dashed, domain =-4:4] {-2.5*\x^3/16+2.5*\x};
  \end{axis}
    \end{tikzpicture}

    \caption{\textbf{Sample interaction functions.} Two cases of coupling functions that we consider. Case 1 (red, dashed) is an odd, $2\pi$-periodic function with a continuous derivative, no zeros in between $0$ and $\pi$, 
    and has a positive slope at 0. Case 2 (blue, solid) is similar to case 1 but has a zero of order 1 in between $0$ and $\pi$.}
    \label{fig:circleofsols}
\end{figure}

\subsection{Identical Oscillators}
We assume that oscillators frequencies are drawn from a known frequency distribution $g(\omega)$. For simplicity we first consider the case of identical oscillators, i.e., we set the distribution to be $g(\omega) = \delta(\omega-\omega_0)$, so the system becomes
\begin{align} \label{eq:sys_ident}
     \dot{\theta}_i = \omega_0 + \dfrac{K}{N} \sum_{j=1}^N f(\theta_j-\theta_i),\quad i=1,\ldots,N\;.
\end{align}

\subsection{Bimodal equilibria}
We assume that the number of oscillators is large, $N \gg 1$, and we look for bimodal equilibria by making the ansatz of an oscillator phase distribution $h(\theta) = x \delta(\theta-\theta_1) + (1-x) \delta(\theta-\theta_2)$, where $0 < x < 1$ describes the fraction in cluster 1.  Note that this constitutes an explicit restriction to a bimodal manifold within the broader space of all possible oscillator phase distributions. Then system \eqref{eq:sys_ident} can be reduced to two coupled ordinary differential equations (ODEs):
\begin{align}
     \dot{\theta}_1 &= \omega_0 + \dfrac{K}{N} \left(\sum_{i=1}^{xN} f(\theta_1-\theta_1)+ \sum_{i=xN+1}^{N}f(\theta_2-\theta_1)\right) \nonumber \\
    &= \omega_0 + K(1-x) f(\theta_2-\theta_1) \label{eq:theta1dot} \\
    \dot{\theta}_2 &= \omega_0 + \dfrac{K}{N} \left(\sum_{i=1}^{xN} f(\theta_1-\theta_2)+ \sum_{i=xN+1}^{N}f(\theta_2-\theta_2)\right) \nonumber  \\
    &=  \omega_0 - K x f(\theta_2-\theta_1) \label{eq:theta2dot}~.
\end{align}
We define a new phase-difference variable $\psi = \theta_2 -\theta_1$ and write its dynamical system by subtracting Eq.~\eqref{eq:theta1dot} from Eq.~\eqref{eq:theta2dot}:
\begin{align} \label{eq:psidot}
   \dot{\psi} = -K f(\psi)~.
\end{align}
We observe that the fixed points of the system for $\psi$ are fully determined by the zeros of $f(\psi)$. From the assumptions above $f(\psi)$ must have zeros at $\psi = 0$ and $\psi = \pi$. Furthermore, if conditions (\ref{ZeroZeroCond}--\ref{PeriodicFun}) hold and $f(\psi)$ has no other zeros (as in the case of the red dashed curve from Fig.~\ref{fig:circleofsols}), then it is implied that $f'(\pi) \leq 0$. Hence, within the bimodal manifold, the fixed point at $\psi =\pi$ should be stable with $\psi = 0$ being unstable.  $\psi = \pi$ corresponds to a bimodal equilibrium with two clusters of oscillators separated by 180$^\circ$ of phase.

If additional roots of $f(\psi)$ exist between $0$ and $\pi$, these will also correspond to bimodal fixed points with alternating stability (again restricted to the bimodal manifold). We focus on the cases where there are no other fixed points or there is exactly one other fixed point $\psi_0$ in $(0,\pi)$; other cases are similarly tractable.  Figure $\ref{fig:circleofsols}$ illustrates the typical general shapes of the interaction functions that we consider.

\subsubsection{Stability of bimodal equilibrium}
To investigate the broader stability of solutions to perturbations outside the bimodal manifold, we consider the perturbation of a single oscillator by a small amount $\epsilon$. Because $N \gg 1$, we approximate the dynamics of the two clusters as unaffected by this perturbation. We examine the evolution of distance between the perturbed oscillator and the group from which it was perturbed, $\epsilon(t)$, to evaluate whether the system returns to its initial state. 

 For convenience, we move into a rotating frame by redefining $\theta_i \rightarrow \theta_i + \omega_0 t$, which is equivalent to setting $\omega_0 =0 $. Without loss of generality we choose oscillator index $N$ from the $\theta_2$ cluster for the perturbation and assume $\theta_1 = 0$, and thus $\theta_2 = \psi_0 \leq \pi $ (assuming for now that our interaction function has only one or zero fixed points in $(0,\pi)$). Then $\theta_N = \theta_2-\epsilon = \psi_0-\epsilon$, and
%
\begin{align*}
    \dot{\epsilon} &=  -\frac{K}{N} \left[\sum_{i=1}^{xN} f(\theta_1-\psi_0+\epsilon)+ \sum_{i=xN+1}^{N-1}f(\theta_2-\psi_0+\epsilon)\right]  \\
    &= - K x f(-\psi_0+\epsilon) - K(1-x)f(\epsilon) \; . 
\end{align*}
We expand the functions $f$ in a Taylor series to linear order:
\begin{equation*}
    \dot{\epsilon}  \approx - \epsilon K \left[   x f'(\psi_0)  +(1-x) f'(0)  \right] \;.
\end{equation*}
%
Assuming that $K<0$ (repulsive coupling, our case of interest in this manuscript), this implies stability if and only if
\begin{equation}
     x f'(\psi_0)  +(1-x) f'(0)  < 0  \;. \label{eq:Nperb}
\end{equation}

A nearly identical calculation starting with the perturbation of a single oscillator from the $\theta_1$ (zero phase) cluster leads to a similar equation,
\begin{equation}
      (1-x) f'(\psi_0)  + x f'(0) < 0 \;.\label{eq:zeroperb}
\end{equation}
Since Eqs.~\eqref{eq:Nperb} and \eqref{eq:zeroperb} must be simultaneously satisfied for stability of the full bimodal distribution, the following inequality must hold:
\begin{equation}
    f'(0) < (1-x) [f'(0)-f'(\psi_0)] < -f'(\psi_0) \;. \label{PreBiModCond}
\end{equation}
Interestingly, this implies that the slope of the interaction function $f(\psi)$ must be steeper at $\psi = \psi_0$ compared to $\psi = 0$ if the bimodal state is to be stable. We can also compute explicit bounds on the proportion of the oscillators in each group by isolating fraction $x$ in inequality $(\ref{PreBiModCond})$ :
\begin{equation}
    \dfrac{f'(0)}{f'(0)-f'(\psi_0)} < x < \dfrac{-f'(\psi_0)}{f'(0)-f'(\psi_0)}.
    \label{eq:BiModCond}
\end{equation}

\begin{figure}[t!]
    \centering 
    \begin{tikzpicture}
    \begin{axis}[width = \columnwidth, height = 0.75\columnwidth,grid=both,xmin =-4.01,ymin=-4.01, xmax =4.01, ymax =4.01, xtick={-4,-2,0,2,4}, xticklabels = {$-\pi$, $-\frac{\pi}{2}$,$0$, $\frac{\pi}{2}$,$\pi$},ytick={-4,-2,0,2,4}, yticklabels = {$-6$, $-3$,$0$, $3$,$6$}]
    \node at (-0.5,3.75)  {$f(\psi)$};
    \node at (3.50,-0.5) {$\psi$};
    \draw[ultra thick] (-4,0) -- (4,0);
    \draw[ultra thick] (0,-4) -- (0,4);
  \addplot[samples=100,blue,ultra thick, domain =-4:4]{(\x+4)*\x*(\x-4)*(\x*\x-9)/16/9)};
    \addplot[samples=100,red,ultra thick, dashed, domain =-4:4] {(\x+4)*\x*(\x-4)*(\x*\x-6)/16/6)};
        \addplot[samples=100,black,ultra thick, dotted, domain =-4:4] {(\x+4)*\x*(\x-4)*(\x*\x-3)/16/3)};
  \end{axis}
    \end{tikzpicture}
    %
     \caption{\textbf{Concrete interaction function.} The interaction function defined in Eq.~\eqref{eq:example1} plotted for several different values of $a$: $\sqrt{3}\pi/4$ (black, dotted), $\sqrt{6}\pi/4$ (red, dashed), and $3\pi/4$ (blue, solid).  As the value of $|a|$ approaches $\pi$ the slope at zero stays fixed with slope 1 and the slope at $\pm a$ decreases in magnitude. This relation between $a$ and the slope values at $\pm  a$, combined with Eq.~\eqref{eq:BiModCond} leads to the threshold for bimodality given by Eq.~\eqref{eq:acrit}. }
    \label{fig:interactionfunc}
\end{figure}
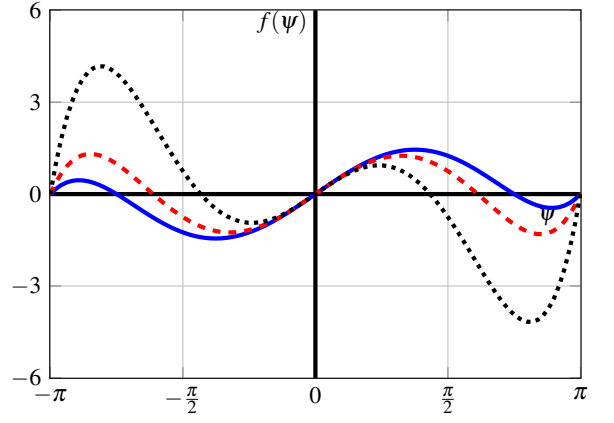

\section{Concrete example}

As a concrete example, we consider a simple class of interaction functions
\begin{equation}
    f(u;a) = \frac{1}{\pi^2 a^2}u\left(\pi^2-u^2\right)\left(a^2-u^2\right).
    \label{eq:example1}
\end{equation}

These functions have roots on $(-\pi, \pi]$ at $0$, $\pi$, and $\pm a$, and satisfy all the conditions set forth earlier in section \ref{sec:modelintro}. As long as $0 < |a| < \pi$ there are three roots in $0 \leq u \leq \pi$, and one can check that $f'(0) = 1$ for all choices of $a$ (see Fig.~$\ref{fig:interactionfunc}$ for example plots).  For inequality \eqref{eq:BiModCond} to be satisfiable, we require
\begin{equation*}
    \frac{\pi^2}{3\pi^2-2a^2}  < \frac{2\pi^2-2a^2}{3\pi^2-2a^2} \;,
\end{equation*}
which reduces to
\begin{equation}
    |a| <  \pi/\sqrt{2} \equiv \acrit \;.
    \label{eq:acrit}
\end{equation}

\begin{figure}[t!]
    \centering
    \includegraphics[width=0.99\columnwidth]{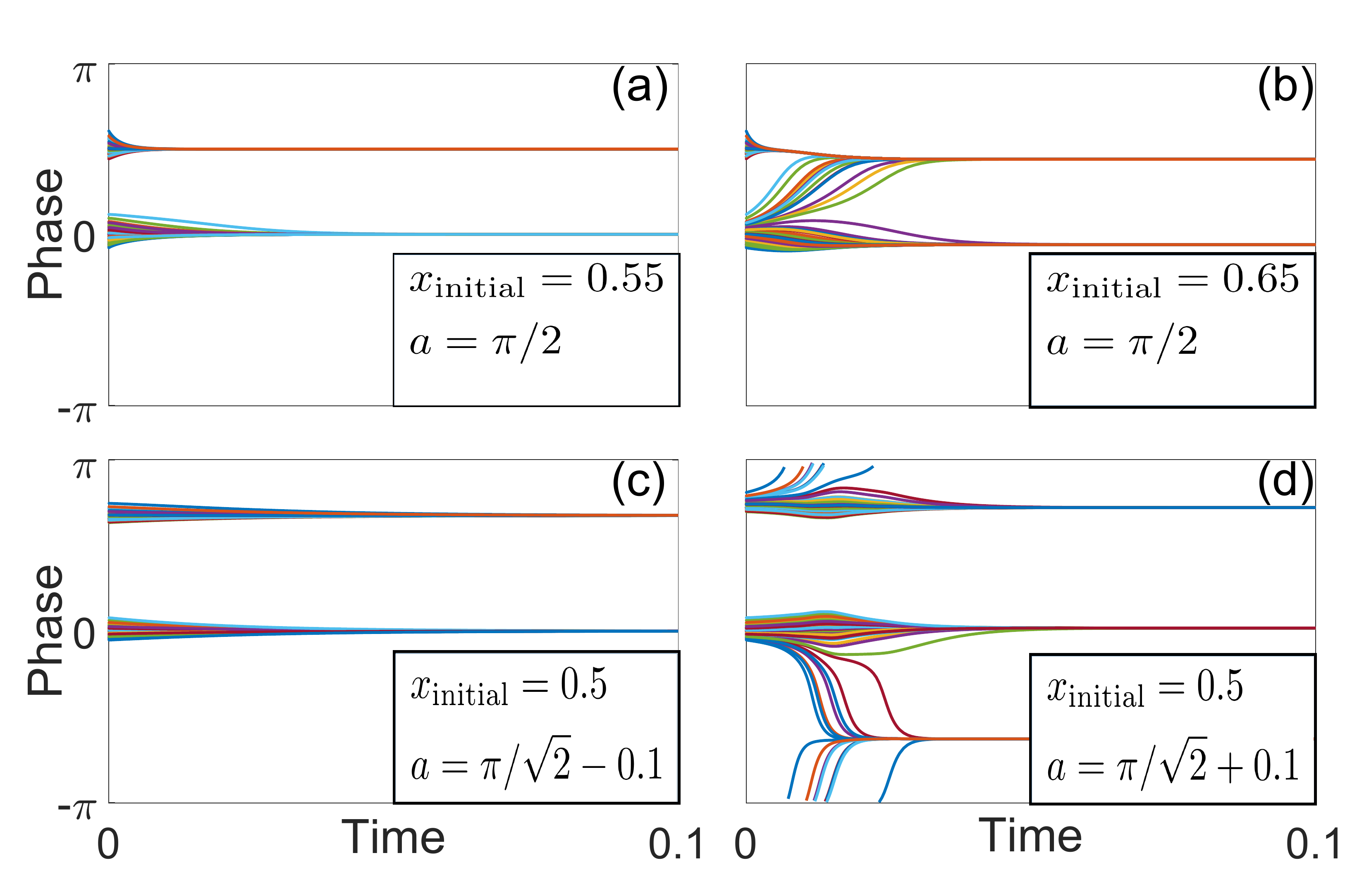}
    \caption{\textbf{Numerical experiments with identical oscillators.} Using example from Eq.~\eqref{eq:example1}, top two panels show test for stability range of fractionation $x$ from Eq.~\eqref{eq:BiModCond}; bottom two panels show test for critical parameter $\acrit$ from Eq.~\eqref{eq:acrit}. (a) When initial fractionation is in the stable range (here $0.4 < \xinit = 0.55 < 0.6$) perturbations shrink and the solution returns to its initial state. (b) When initial fractionation is outside stable band (here $\xinit = 0.65 > 0.6$) perturbations grow for some oscillators until system evolves to a different fractionation state.  (c) When $\xinit = 1/2$ and $a < \acrit$, perturbations shrink and the solution returns to its initial state. (d) When $\xinit = 1/2$ and $a > \acrit$, perturbations grow and the system moves away from the unstable bimodal state until it reaches a new trimodal equilibrium. } 
    \label{fig:numericaltests}
\end{figure}

We note that symmetry of the roots allows us to consider positive $a$ without loss of generality. Figure \ref{fig:numericaltests} shows the results of numerical experiments where we test this predicted stability threshold.  In each panel, Eq.~\eqref{eq:sys_ident} is implemented with the interaction function from Eq.~\eqref{eq:example1}.  We initialize $xN$ oscillators at $\theta_1 = 0$ and $(1-x)N$ at $\theta_2 =a$, then add a small random perturbation $\xi_i$ to each oscillator's initial phase, where $\xi_i$ is drawn from the normal distribution $\mathcal{N}(0,\delta^2)$, with $\delta = 0.1$ used in Fig.~\ref{fig:numericaltests}. We numerically integrate the system using a 4th/5th order Runge-Kutta scheme and consider evidence for stability if it approaches the unperturbed state, i.e. $\psi= \theta_2 - \theta_1 \rightarrow a$ with $\xfinal = \xinit$.  We note that in these experiments we set coupling strength $K = -1000$.

In panels (a) and (b), we use $N = 100$ oscillators, $\omega_0 = 0$, and set $a = \pi/2$, consistent with the stability threshold from Eq.~\eqref{eq:acrit}, $a < a_{\textrm{crit}}=\pi/\sqrt{2}$. The stable band of fractionation according to inequality \eqref{eq:BiModCond} is then $2/5 < x < 3/5$. In panel (a), we set $\xinit = 0.55$, below the band's upper bound; in panel (b), we set $\xinit = 0.65$, above the band's upper bound.  As expected, the bimodal equilibrium appears stable in panel (a), but unstable in panel (b), where eleven oscillators move between clusters to establish a different equilibrium within the stable fractionation band ($2/5 < x_\text{final}=0.54 < 3/5$).



In panels (c) and (d), we again use $N = 100$ oscillators and $\omega_0 = 0$, but here we examine the predicted stability threshold $\acrit = \pi/\sqrt{2}$ from Eq.~\eqref{eq:acrit}.  We expect the bimodal state with $\psi^*=a$ to be unstable for all positive $a > \acrit$ (but note that this state ceases to exist when $a > \pi$).  We set $\xinit = 1/2$ since this is within the fractionation stability band from inequality \eqref{eq:BiModCond} for all $a <  \acrit$. In panel (c), we set $a = \acrit - 0.1$, just below the threshold for stability; in panel (d), we set $a = \acrit + 0.1$, just barely in the unstable domain.  As expected, the bimodal equilibrium again appears stable in panel (c), but it appears unstable in panel (d).  Since no fractionation $x$ will lead to a stable bimodal equilibrium, the system must move to an entirely different state, and it appears to converge to a trimodal distribution of oscillator phases. 

We are able to understand why the system converges to a trimodal state by performing a similar analysis for the stability of three-cluster, or trimodal, oscillator distributions. One can show that a necessary condition for stability is:
\begin{align}
     f'(0) <  &-\left[(x+y) f'(\psi_1) +(y+z) f'(\psi_2) \right. \nonumber \\
     &\left. +(x+z) f'(\psi_1+\psi_2)\right]
\end{align}
where $\psi_i$ is the angle separating clusters at $\theta_i$ and $\theta_{i+1}$ ($\theta_4$ identified with $\theta_1$), and $x$, $y$, and $z$ are the fractionations of the three clusters at $\theta_1$, $\theta_2$, and $\theta_3$ respectively. With equal spacing between the clusters $\psi_1 = \psi_2 = 2\pi -\psi_1-\psi_2= 2\pi/3$, the necessary condition simplifies to
\[
    f'(0) <  -2 f'(\frac{2\pi}{3})\;.
\]
For the example function shown in Eq.~\eqref{eq:example1} this is 
\[
    a < \frac{2}{3}\sqrt{\frac{14}{3}} \pi \equiv a_{\text{tricrit}} \approx 1.44 \pi\;.
\]
This implies that a trimodal state remains stable for all $a < \pi$. It stably coexists with the bimodal state for $a < \pi/\sqrt{2}$, and may coexist with other multimodal states for $\pi/\sqrt{2} < a < \pi$.
In general different multimodal states may stably coexist over various parameter ranges. More details of the analysis for trimodality  can be found in the appendix.



\section{Generalization to asymmetric interaction functions}

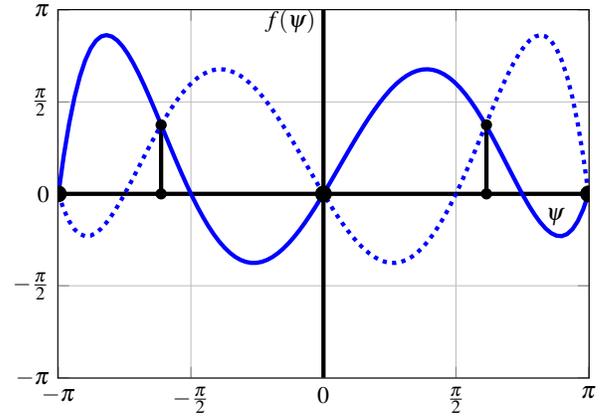
\begin{figure}[t!]
    \centering 
    \begin{tikzpicture}
    \begin{axis}[width = \columnwidth, height = 0.75\columnwidth,grid=both,xmin =-4.01,ymin=-4.01, xmax =4.01, ymax =4.01, xtick={-4,-2,0,2,4}, xticklabels = {$-\pi$, $-\frac{\pi}{2}$,$0$, $\frac{\pi}{2}$,$\pi$},ytick={-4,-2,0,2,4}, yticklabels = {$-\pi$, $-\frac{\pi}{2}$,$0$, $\frac{\pi}{2}$,$\pi$}]
    \node at (-0.5,3.75)  {$f(\psi)$};
    \node at (3.5,-0.50) {$\psi$};
    \node[fill =black,shape =circle,scale=0.5] at (-2.45,0 ) {};
    \node[fill =black,shape =circle,scale=0.5] at (2.46,0 ) {};
     \draw[ultra thick] (-2.45,0) -- (-2.45,1.50);
    \draw[ultra thick] (2.46,0) -- (2.46,1.50);
    \draw[ultra thick] (-4,0) -- (4,0);
     \draw[ultra thick] (0,-4) -- (0,4);
  \addplot[samples=100,blue,ultra thick, domain =-4:4]{(\x+4)*\x*(\x-4)*(\x-3)*(\x+2)/40)};
      \addplot[samples=100,blue,ultra thick, dotted,domain =-4:4]{-(\x+4)*\x*(\x-4)*(\x+3)*(\x-2)/40)};
        \node[fill =black,shape =circle,scale=0.5] at (-2.45,1.50 ) {};
        \node[fill =black,opacity=1,shape =circle,scale=0.5] at (2.46,1.50 ) {};
      \node[fill =black,opacity=1,shape =circle,scale=0.75] at (0,0 ) {};
    \node[fill =black,shape =circle,scale=0.75] at (-4,0 ) {};
    \node[fill =black,shape =circle,scale=0.75] at (4,0 ) {};
  \end{axis}
    \end{tikzpicture}

    \caption{\textbf{Sample asymmetric interaction function.} This function (solid blue curve) does not satisfy $f(\psi) = -f(-\psi)$. Existence of bimodal equilibria requires that it intersect its mirror reflection (dotted blue curve) or a scaled version of it (see Eq.~\eqref{eq:asymFPs}).  The fixed points of the system for $x=1/2$ are marked by black dots.}
    \label{fig:NonAntiSymPlot}
\end{figure}

We can relax assumption \eqref{Oddfun} of an antisymmetric coupling function and still find stability boundaries for multimodal states.  In place of Eq.~$\eqref{eq:psidot}$ (which used oddness of the coupling function), we find instead
\begin{equation}
    \dot{\psi} =   K x f(-\psi) - K (1-x) f(\psi) \;.
\end{equation}
Clearly $\psi^*=0$ and $\psi^*=\pi$ both remain fixed points.  Other fixed points exist if
\begin{equation} \label{eq:asymFPs}
    x f(-\psi^*) = (1-x) f(\psi^*)
\end{equation}
has a solution on $-\pi < \psi^* \leq \pi$.
Figure \ref{fig:NonAntiSymPlot} shows an example of an asymmetric interaction function.  Geometrically this condition can be understood as identifying intersections of $f(\psi)$ and its reflection $f(-\psi)$ when $x=1/2$ (or scaled versions when $x \neq 1/2$).  Once multimodal fixed points are identified, stability analysis is analogous to that presented earlier.

\section{Generalization to non-identical oscillators}

We argue that real-world bimodal or multimodal distributions may result from similar dynamics to those presented in this paper. Of course, heterogeneity is inevitable in most real-world systems, yet we have focused thus far on the case of identical oscillators.  While we leave the more general analysis for future work, we have conducted numerical experiments that appear to show that the predicted behavior occurs even in the presence of oscillator heterogeneity.

\begin{figure}[t!]
    \centering
    \includegraphics[width = 0.99\columnwidth]{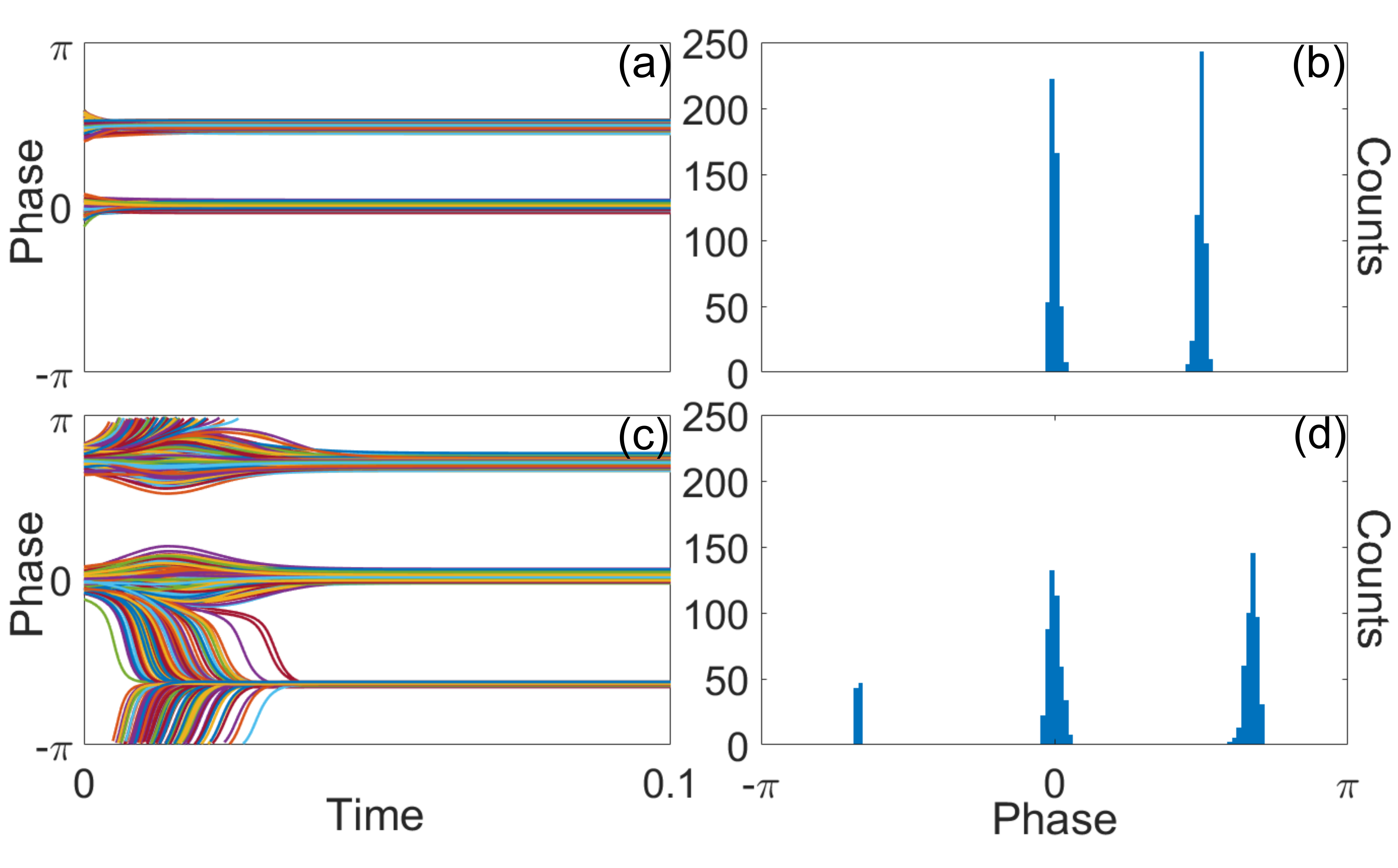}
    \caption{\textbf{Numerical experiments with heterogeneous oscillators.} Here, $N = 1000$ and oscillators' frequencies are drawn from the distribution $\mathcal{N}(0,100)$ and the perturbations, $\xi_i$, $i= 1\ldots N$, are drawn from $\mathcal{N}(0,0.01)$. Using example from Eq.~$\eqref{eq:example1}$, panels (a) and (b) show the results for $\xinit = 1/2$ and $a = \pi/2 < \acrit$ (compare to Fig.~\ref{fig:numericaltests}(a)). 
    Panels (c) and (d) show the results for $\xinit = 1/2$ and $a = \pi/\sqrt{2}+0.1 > \acrit$ (compare to Fig.~\ref{fig:numericaltests}(d)).
    }
    \label{fig:BimodNonUni}
\end{figure}

Again using the same example interaction function from Eq.~\eqref{eq:example1}, we now draw frequencies, $\omega_i$, from a normal distribution $\mathcal{N}(0,\sigma^2)$ and set the initial phases of the oscillators to $\theta_i = \xi_i$ (fraction $x$) or $\theta_i = a + \xi_i$ (fraction $1-x$), where $\xi_i$ is a small perturbation draw from the distribution $\mathcal{N}(0,\delta^2)$.  Figure \ref{fig:BimodNonUni} shows the results of perturbation experiments analogous to those presented in Fig.~\ref{fig:numericaltests}, with analogous results except that the final phase distributions have phases that cluster about the modes rather than all converging to them precisely (right panels show histograms of final states). 

In Fig.~\ref{fig:BimodNonUni} panels (a) and (b), we use $N=1000$ oscillators and set $a = \pi/2 <a_{\text{crit}}$ and $x_{\text{initial}} =1/2$.  Even with perturbed initial phases and heterogeneous natural frequencies, the oscillators still remain in the bimodal state as predicted for $a<a_{\text{crit}}$.  Specifically, panel (b) shows that the steady state distribution of oscillators has finite-width clustering about the fixed point positions predicted from the identical-oscillator case. In panels (c) and (d), since $a =\pi/\sqrt{2}+0.1 > a_{\text{crit}}$, the bimodal state breaks down (consistent with the prediction of the identical-oscillator theory) and the system appears to converge to a trimodal equilibrium with three finite-width clusters.



\section{Discussion}

Coupled oscillators are an excellent testbed for models of synchronization or clustering.  Even though real-world variables (e.g., sediment grain size\cite{sun2004}, salmon body size\cite{thorpe1977,damsgaard2019proactive}, human communication frequency\cite{wu2010}, dopamine receptor density\cite{seeman1984}, neutron star mass\cite{schwab2010}, galaxy color\cite{baldry2004}, gamma ray burst duration\cite{mao1994}, tree height\cite{eichhorn2010}, animal ornament size\cite{clifton2016}) may not be oscillatory or confined to a periodic domain, bimodality may emerge for qualitatively similar reasons.  In our model, the coupling of one unit's dynamical behavior to that of others is key to the phenomenon. 

For clarity of presentation we have focused on a single example of interaction function (Eq.~\eqref{eq:example1}), but evaluation of two other classes of interaction functions (triangle waves and antisymmetrized von-Mises kernels) also supports our analytical results---see supplementary material for details.  In supplementary material, we also present further results regarding dependence of bimodal equilibria on coupling strength $K$, as well as some numerical evidence regarding sizes of basins of attraction; each of these topics merits further in-depth study.  
%
The analysis we present here focuses exclusively on the case of all-to-all coupling; we leave further investigation of the impact of network structure for future work.

For real-world scenarios where bimodality or multimodality is of interest, the interaction function may not be known exactly.  Nevertheless, we expect that it will often be possible to assess whether the conditions expressed in Eqns.~\eqref{allconditions} hold in a particular case.  It also seems plausible that functions describing real-world interactions between coupled systems will have no more than a handful of roots, making bimodality and trimodality likely outcomes when repulsive or inhibitory coupling is imposed. 

One particularly important case occurs when the interaction function has only roots at zero and $\pi$, with the root at zero having larger or equal magnitude slope.  That is the case in the standard Kuramoto model with sinusoidal coupling.  In such a case we expect that the incoherent splay state will be stable.  In general, the splay state should be stable when the tendency to cluster (due to long-distance interactions) cannot overcome the oscillators' locally repulsive interactions.




\section{Conclusions }

We have shown that, when coupling is repulsive, multi-modality of the oscillator distribution can be a stable configuration for a wide range of interaction functions. We showed that bimodality can be expected under repulsive coupling when the slope of the interaction function at the origin is shallower than at the other root(s).  We performed numerical experiments for both identical and nonidentical oscillators and observed results consistent with theory. 

This demonstration that repulsive coupling can produce clustering under reasonable assumptions about the interaction dynamics is important as repulsive coupling is present in many natural systems. Hence, the theory we present in this paper provides an argument as to why one might expect multi-modality instead of unimodality or incoherence in systems known to have repulsive coupling.

\section*{Supplementary Material}

See accompanying supplementary material for numerical experiments using a selection of additional interaction functions, for discussion of basins of attraction for different states, and for discussion of coupling strength dependency.


{
\appendix*
\section{Trimodal equilibria}
We again consider a function $f(u)$ that satisfies conditions \eqref{ZeroZeroCond}--\eqref{PeriodicFun}. We look for solutions with oscillators distributed according to $h(\theta) = x \delta(\theta-\theta_1)+y \delta(\theta-\theta_2) + (1-x-y) \delta(\theta-\theta_3)$, where $x, y> 0$, $x+y <1$ so that the oscillators will be in three clusters at $\theta_1$, $\theta_2$, and $\theta_3$ (we again assume that the natural frequencies are identical):
\begin{align}
    \dot{\theta}_1
    &= \omega_0 +  K \left( y f(\theta_2-\theta_1)+ z f(\theta_3-\theta_1) \right)  \\
     \dot{\theta}_2
    &= \omega_0 +  K \left( -x f(\theta_2-\theta_1)+ z f(\theta_3-\theta_2) \right)  \\
     \dot{\theta}_3
    &= \omega_0 -  K \left( x f(\theta_3-\theta_1)+ y f(\theta_3-\theta_2) \right)  \;.
\end{align}
Here, $z = 1-x-y$. We define two variables $\psi_1 = \theta_2 -\theta_1$ and $\psi_2 = \theta_3 -\theta_2$, so that the system reduces to
\begin{align}
     \dot{\psi}_1 &= -K \left( z \left[ f(\psi_2+\psi_1) -f(\psi_2)  \right] +(x+y) f(\psi_1)\right)  \\ 
  \dot{\psi}_2  &=   -K \left( x \left[ f(\psi_2+\psi_1) -f(\psi_1)  \right] +(y+z) f(\psi_2)\right) \;.
\end{align}
We set $\dot{\psi}_i = 0$, $i = 1, 2$ and arrive at the following system of equations:
\begin{align}
    f(\psi_2) &= \dfrac{xf(\psi_1)}{z} \label{TriModEq1} \\
     f(\psi_2+\psi_1) &= \dfrac{-yf(\psi_1)}{z} \label{TriModEq2}.
\end{align}
To set bounds on the fractionation of the clusters, we assume that there exists points $\psi_1,\psi_2 \in (-\pi,\pi)$ such that Eqns.~\eqref{TriModEq1} and \eqref{TriModEq2} are satisfied. Additionally, we put our system of coupled oscillators into a rotating frame so that $\theta_i \rightarrow \theta_i +\omega_0 t$. In the rotating frame, we set $\theta_1=0$, $\theta_2 =\psi_1$, and $\theta_3 = \psi_1 +\psi_2 -2\pi$. As before we perturb an oscillator from one of the three groups. We do this for all three groups and get a system of inequalities
\begin{subequations}
\begin{align}
 x f'(0) +y f'(\psi_1) +z f'(\psi_1+\psi_2) &<  0\\ 
 y f'(0) +x f'(\psi_1) +z f'(\psi_2) &<  0 \\
  z f'(0) + y f'(\psi_2) +x f'(\psi_1+\psi_2) &<  0 \; .
\end{align}
\end{subequations}
All these must be simultaneously satisfied for stability of a trimodal state.  Adding, we find
\begin{align}
     f'(0) <  &-\left[(x+y) f'(\psi_1) +(y+z) f'(\psi_2) \right. \nonumber \\
     &\left. +(x+z) f'(\psi_1+\psi_2)\right] \;. \label{gentrimodNecCond}
\end{align}
This states that the weighted sum of the slopes of the coupling function at $\psi = \psi_i$, where the weights are the proportions for the groups separated by $\psi_i$, is greater in magnitude than the slope at the origin. This condition reduces to
\begin{align}
    f'(0) < -2 f'\left(\dfrac{2\pi}{3}\right) \label{StabTriMod}
\end{align}
if $\psi_1 = \psi_2 =2\pi -\psi_1-\psi_2 =2\pi/3$. As an example, we return to the class of interaction functions that we introduced in Section II. We relax the assumption that $|a| < \pi$ and consider the case when $\psi_1 = \psi_2 =2\pi -\psi_1-\psi_2$. To satisfy inequality $(\ref{StabTriMod})$, this means that
\begin{align}
    1 < \frac{56\pi^2+54a^2}{81 a^2} \;,
\end{align}
which reduces to 
\begin{align}
    |a| < \frac{2}{3}\sqrt{\frac{14}{3}}\pi \equiv \atricrit \approx 1.44 \pi \;. \label{eq:trimodthresh}
\end{align}
Figure \ref{fig:TriModCases} shows the results of a numerical experiment where we test this threshold. In both panels we use $N = 99$, $x=y=z=1/3$, $\psi_1 = \psi_2 =2\pi -\psi_1-\psi_2 =2\pi/3$, and set $\omega_0=0$. We expect the trimodal state to be unstable for $a > a_{\text{tricrit}}$. In panel (a) we set $a = a_{\text{tricrit}}+0.1$ and perturb the oscillators by amount $\xi_i$, with values drawn from the distribution $\mathcal{N}(0,0.01)$ . We can see that this perturbation leads to the system leaving the trimodal state and going to a bimodal state with $180^\circ$ phase difference. 

One might be interested in  why the bimodal state is stable in panel (a). Since there are only zeros at $\psi = 0$ and $\psi = \pi$, one may check the stability by evaluating the derivative of $f(\psi)$ at these points. One can show that if
\begin{equation}
|a|  > \sqrt{2}\pi \label{eq:antiphasecond}
\end{equation} 
the $180^\circ$ antiphase state is stable. Thus, when $a = a_{\text{tricrit}}+0.1 > \sqrt{2} \pi$ the trimodal state becomes unstable and perturbations lead to the stable bimodal state. 

For the case, when  $\sqrt{2} \pi < a <a_{\text{tricrit}}  $, both the trimodal state and the bimodal state are stable configurations. Figure \ref{fig:BistabCases} shows the result of the numerical experiment where we place the parameter inside the previously stated interval and outside of the interval. In all panels we use $N=300$, and set $\omega_0 =0$. As before, in all panels we perturb the  oscillators by amount $\xi_i$ from the predicted fixed points, whose values are drawn from the distribution $\mathcal{N}(0,0.01)$. In panels (a) and (b) we set $a =1.43\pi \in (\sqrt{2}\pi,a_{\text{tricrit}})$. In these cases we expect both the bimodal state and the trimodal state to be stable for this value of $a$. In panels (a) and (b), we set the fractionation to be equal in all groups, and we set the spacing between groups to be equal. As expected, we see that the trimodal state and the bimodal state are stable under perturbation.  

In panel (c) we set $a = 1.43\pi-0.1  < \sqrt{2}\pi < a_{\text{crit}} $. As expected, we see that the bimodal state is unstable and the system goes in to trimodal state. Given the proximity of the clusters to $\pm \pi$, we have added black dashed lines that at $\pm \pi$, so that one can see that the difference between the final state and $\pm \pi$.  In panel (d), we set $a = 1.43\pi +0.1 > a_{\text{tricrit}} > \sqrt{2}\pi$. We also observe an expected result, as trimodality appears to be unstable and the system converges to a bimodal equilibrium, which is stable given that $a > \sqrt{2}\pi$.

In summary, we have a necessary condition for the stability of the trimodal equilibrium. Although, this condition is only necessary for stability, not sufficient, numerical experiments seems to point to it being an accurate threshold in examples we have considered. Also, theory and numerical experiments demonstrate that multistability of different multimodal equilibria is possible over parameter space. The theory for the stability of higher modes we leave for future work.

\begin{figure}[t!]
    \centering
    \includegraphics[width= 0.99\columnwidth]{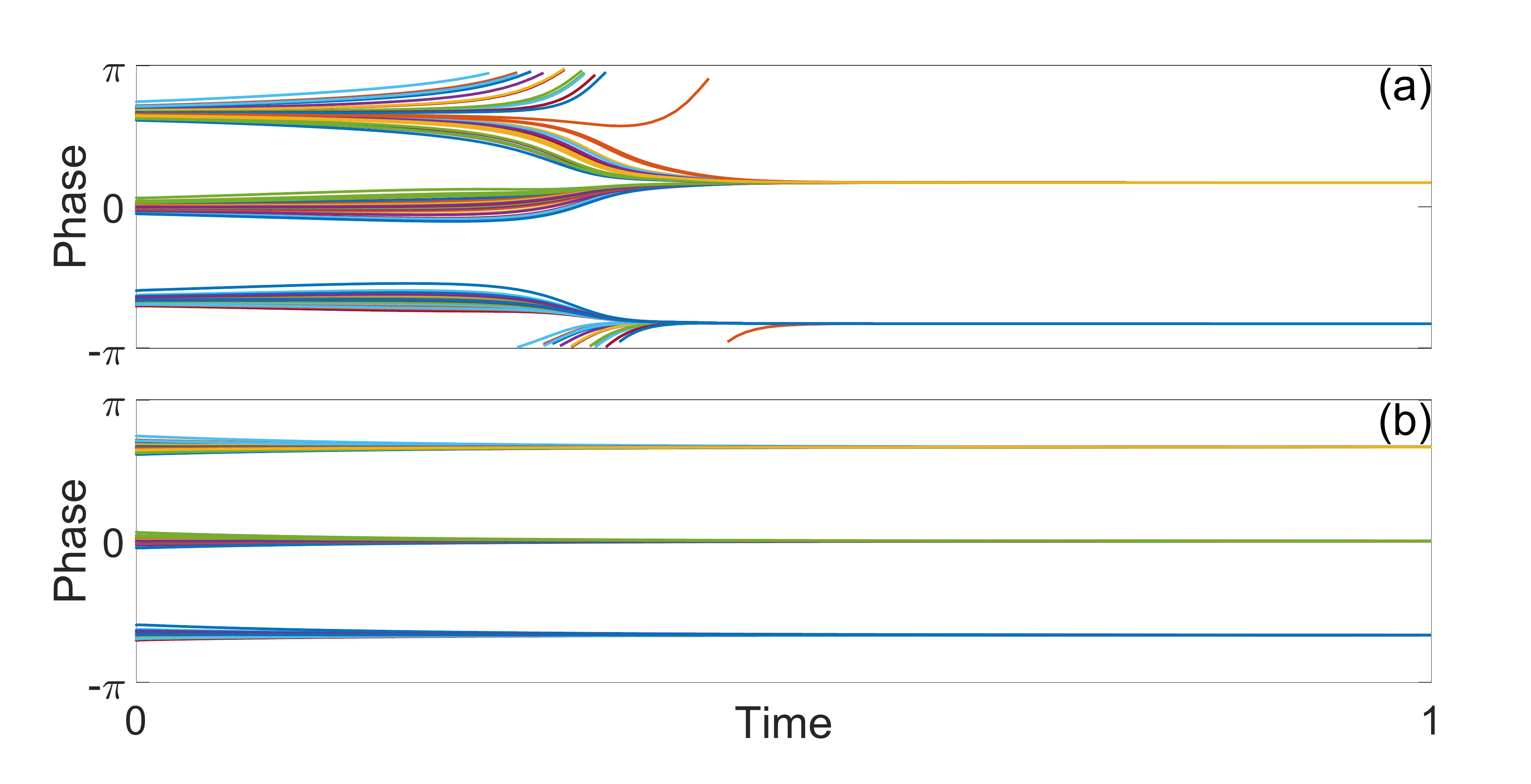}
    \caption{\textbf{Numerical experiments testing the threshold for trimodality.} Panel (a): parameter value is $a  = a_{\text{tricrit}}+0.1$, and the trimodal state appears to be unstable (as expected). Panel (b): parameter va;ie is $a  = a_{\text{tricrit}}-0.1$, and the trimodal state appears to be stable (as expected).  Both panels use the example interaction function from Eq.~\eqref{eq:example1}, and both use equal fractionation ($x=y=z=1/3$) and equal spacing between clusters ($\psi_1= 2\pi/3$) in initial conditions. }
    \label{fig:TriModCases}
\end{figure}

\begin{figure}[t!]
    \centering
    \includegraphics[width= 0.99\columnwidth]{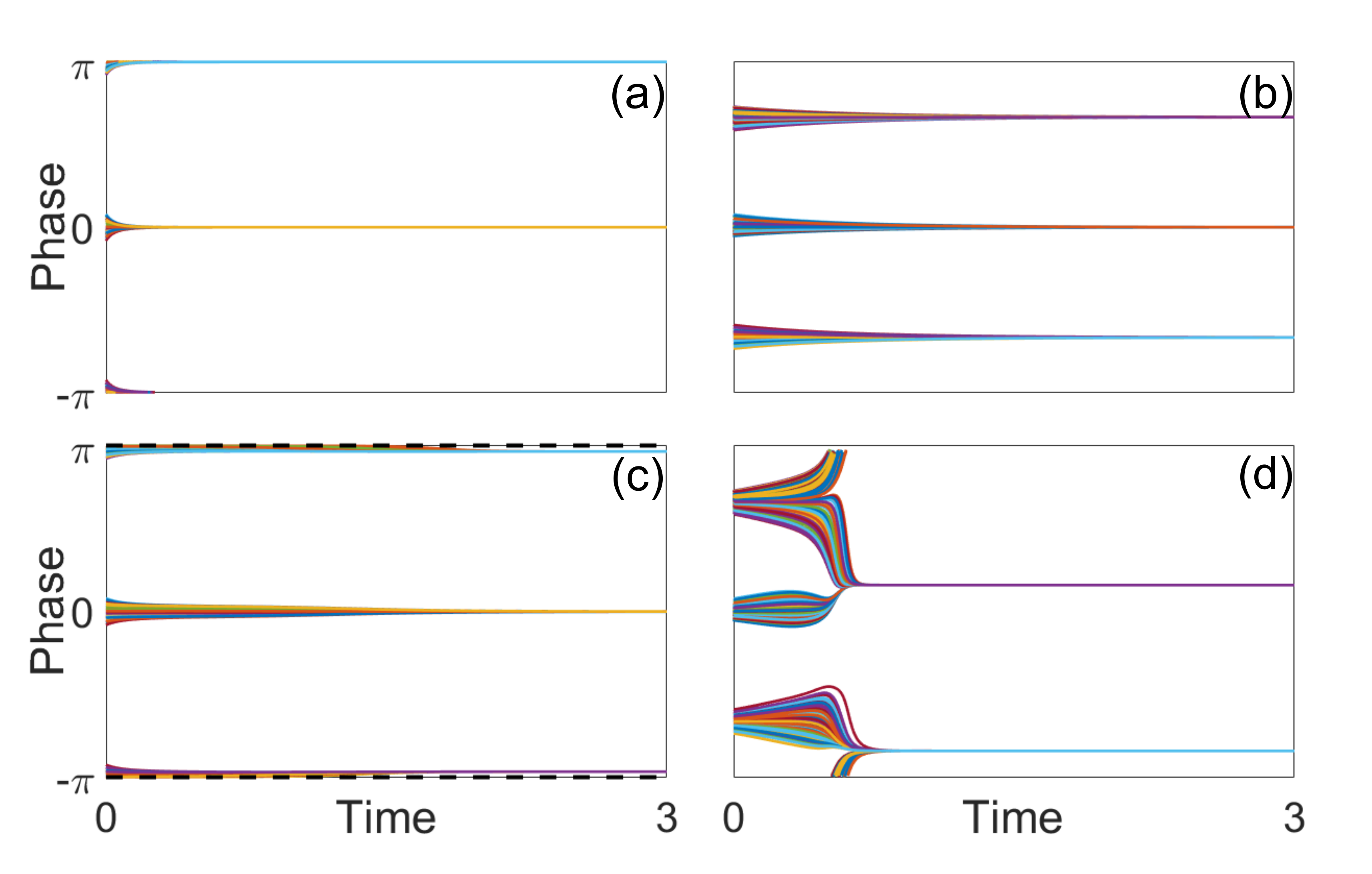}
    \caption{\textbf{Numerical experiments testing bistability.} Panel (a) and (b): we set $a =1.43\pi \in (\sqrt{2}\pi,a_{\text{tricrit}})$ and both the bimodal state and the trimodal state are stable (as predicted). Panel (c): we set $a = 1.43\pi-0.1  < \sqrt{2}\pi < a_{\text{crit}}$ and we see that the bimodal state is unstable (we have added black dashed lines so that one can see that the clusters away from the origin are not at $\pm \pi$). Panel (d): we set $a = 1.43\pi +0.1 > a_{\text{tricrit}} > \sqrt{2}\pi$ and the trimodal state is unstable (as predicted). In all panels $N =300$ and the initial conditions are equally spaced and have equal fractionation with a random perturbation to all the phases of the oscillators.}
    
    \label{fig:BistabCases}
\end{figure}



}
\section*{REFERENCES}
\vspace{-0.6cm}
\bibliographystyle{aipnum4-1}
\bibliography{bimodal}

\clearpage

\setcounter{section}{0}
\setcounter{figure}{0}
\renewcommand\thesection{S\arabic{section}}  
\renewcommand\thefigure{S\arabic{figure}}  
\renewcommand{\theequation}{S\arabic{equation}}

\preprint{AIP/123-QED}

\title{Supplementary Material: A coupled oscillator model for the origin of bimodality and multimodality}

\author{J.D. Johnson}
\author{D.M. Abrams}%
\affiliation{ 
Department of Engineering Sciences and Applied Mathematics, Northwestern University \\ McCormick School of Engineering and Applied Science \\
2145 Sheridan Road
Evanston, IL 60208
}%

\date{\today}
\maketitle

{\LARGE  Supplementary Material: A coupled oscillator model for the origin of bimodality and multimodality}

\begin{figure}[t!]
    \centering 
    \begin{tikzpicture}
    \begin{axis}[width = 0.99\columnwidth, height = 0.7\columnwidth,grid=both,xmin =-4.01,ymin=-4.01, xmax =4.01, ymax =4.01, xtick={-4,-2,0,2,4}, xticklabels = {$-\pi$, $-\frac{\pi}{2}$,$0$, $\frac{\pi}{2}$,$\pi$},ytick={-4,-2,0,2,4}, yticklabels = {-2,-1,0,1,2}]
    \node at (-0.5,3.75)  {$f(\psi)$};
    \node at (3.5,-0.50) {$\psi$};

    \draw[ultra thick] (-4,0) -- (4,0);
     \draw[ultra thick] (0,-4) -- (0,4);
  \addplot[samples=100,blue,ultra thick, domain =-3:3]{4*\x/3};
    \addplot[samples=100,blue,ultra thick, domain =2.3562:4]{4*(-3*\x*.7854+9.4248)/(2.3562)};
        \addplot[samples=100,blue,ultra thick, domain =-4:-2.3562]{4*(-3*\x*.7854-9.4248)/(2.3562)};
      \addplot[samples=100,red,ultra thick, dotted,domain =-4:4]{1.228*sin(deg(\x*.7854))*exp(10*cos(deg(\x*.7854)))/(1769.2)};
      \addplot[samples=100,red,ultra thick,domain =-4:4]{1.228*sin(deg(\x*.7854))*exp(-10*cos(deg(\x*.7854)))/(1769.2)};

  \end{axis}
    \end{tikzpicture}

    \caption{\textbf{Additional interaction functions.} Solid blue curve: triangle wave from Eq.~\eqref{eq:TriWave}; solid red curve: antisymmetrized variant of the von Mises distribution from Eq.~\eqref{eq:vonMises} with $\kappa<0$; dashed red curve: antisymmetrized variant of the von Mises distribution from Eq.~\eqref{eq:vonMises} with $\kappa>0$. Panels (a) and (b) of Fig.~\ref{fig:SI_BimodCases} use the triangle wave. Panels (c) and (d) use the antisymmetrized von Mises function, with positive $\kappa$ (dashed red) in panel (c) and negative $\kappa$ (solid red) in panel (d).  We note that for $\kappa>0$ the slope at the $\pm \pi$ is never steeper when compared to the origin and for $\kappa < 0$ the slope at the origin is never steeper when compared to the slope at $\pm \pi$.}
    \label{fig:AddTestFunc}
\end{figure}
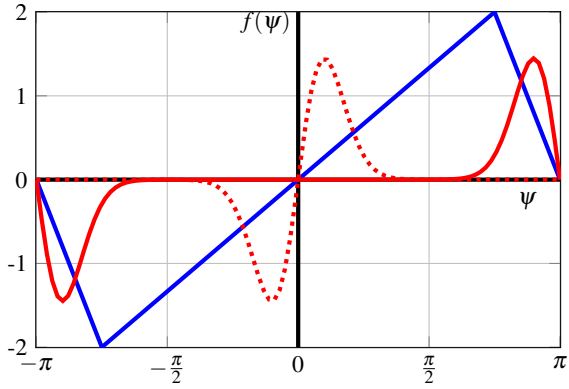

\section{Additional Coupling Functions}

Figure \ref{fig:AddTestFunc} illustrates two additional coupling functions that we examined. We used a variant of the triangle wave (blue, solid) given by the equation
\begin{align}
   f_\text{tri}(u;c)  = \begin{cases} 
      \frac{2u}{c} & |u| < c \\
      \frac{2u}{c-\pi} -\text{sign}(u)(\frac{2\pi}{c-\pi}) & c\leq |u|\leq \pi 
   \end{cases}\;,
   \label{eq:TriWave}
\end{align}
assuming that $0<c<\pi$,
and an antisymmetrized variant of the von Mises distribution (red curves) given by
\begin{align}
  f_\text{vM}(u;\mu,\kappa) = \sin(u-\mu)\frac{e^{\kappa\cos(u-\mu)}}{2\pi I_0(\kappa)} \;.
  \label{eq:vonMises}
\end{align}
\begin{figure}[t!]
    \centering
    \includegraphics[width=0.99\columnwidth]{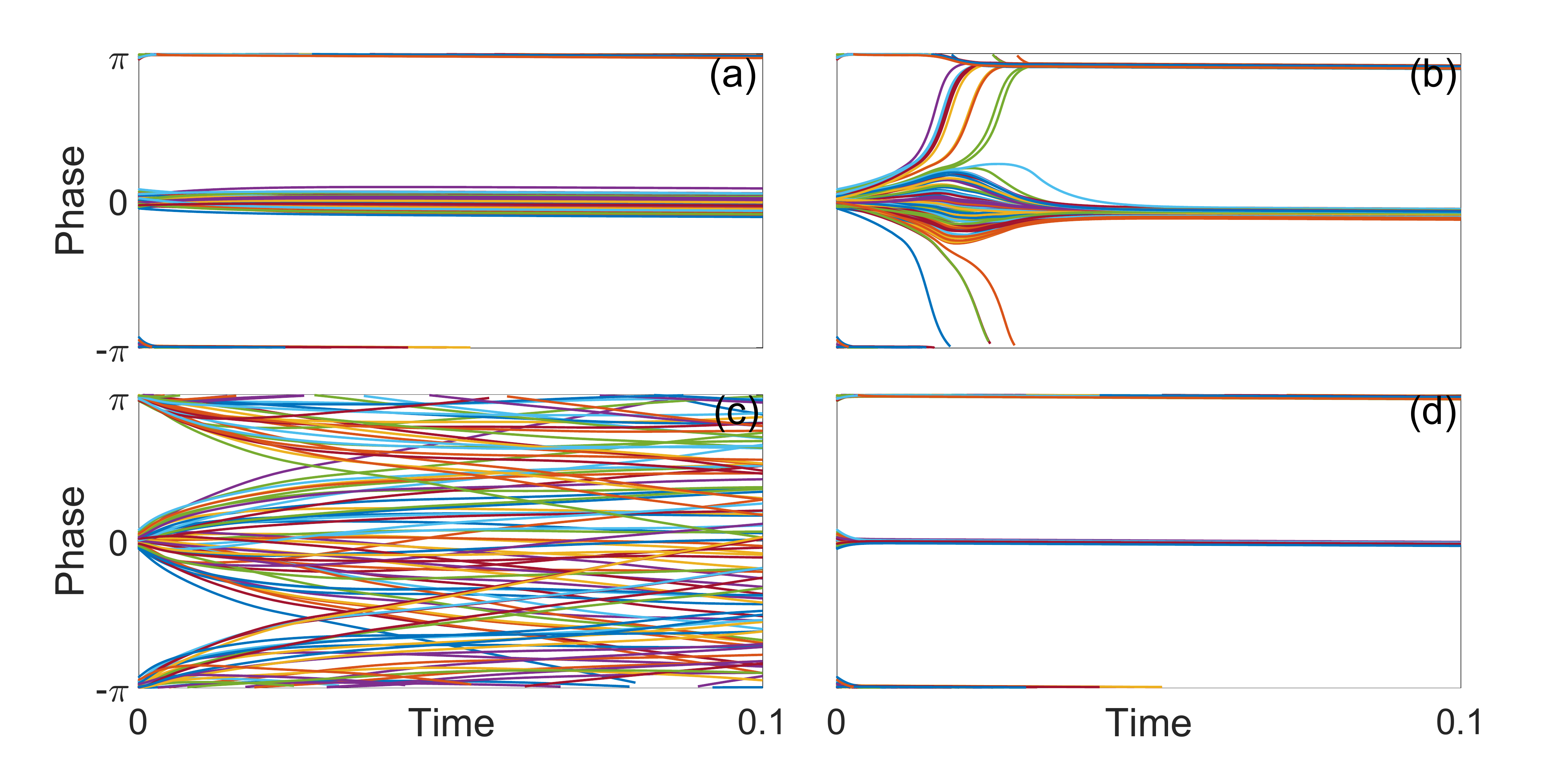}
    \caption{\textbf{Numerical experiments using additional interaction functions.} We test the stability of the bimodal equilibria for alternative coupling functions shown in Fig.~\ref{fig:AddTestFunc}. (a) Triangle wave coupling with initial fractionation in predicted stable range. (b) Triangle wave coupling with initial fractionation outside predicted stable range. (c) Von Mises coupling with $\kappa>0$ (expected to be unstable). (d) Von Mises coupling with $\kappa>0$ (expected to be stable). In all panels $N=100$ and oscillators' natural frequencies are drawn from the distribution $\mathcal{N}(0,100)$. Initial phases are bimodally distributed with modes at $0$ and $\pi$, with perturbations $\xi_i$, $i=1,\ldots, N$, are drawn from $\mathcal{N}(0,0.01)$. 
    }
    \label{fig:SI_BimodCases}
\end{figure}

We numerically probe the stability of the bimodal equilibrium using these interaction functions in Fig.\;\ref{fig:SI_BimodCases}. Here $N = 100$, the  oscillators' frequencies are drawn from a distribution $\mathcal{N} (0,100)$, the phase perturbation, $\xi_i$, is drawn from the distribution $\mathcal{N}(0,0.01)$ and we set $K=-1000$. In panels (a) and (b) we take the triangle wave defined in Eq.~\eqref{eq:TriWave} and set $c = 3\pi/4$; this gives a stable fractionation threshold $1/4<x<3/4$. We test that threshold numerically by setting $x_{\text{initial}} = 7/10 < 3/4$ in panel (a) and  $x_{\text{initial}} = 8/10 > 3/4$ in panel (b). As expected, we see that the fractionation is stable in panel (a) and is unstable in panel (b). 

In panels (c) and (d) we use the antisymmetrized von-Mises function from Eq.~\eqref{eq:vonMises} with $\mu = 0$ and $x_{\text{initial}}=1/2$. In panel (c) we set $\kappa = 10$, and, as expected, we see that the bimodal equilibrium appears unstable; this is because there does not exist a range of $x$ such that Eq.~(10) can be satisfied given that the slope at the origin is far steeper than the slope at the $\pm \pi$.  We note that in  (c) the system appears to tend to the incoherent state. In panel (d) we set $\kappa = -10$ and observe that the bimodal state appears to be stable under perturbation, which is expected given that the slope at the $\pm \pi$ is steeper when compared to the origin.

\begin{figure}[t!]
    \centering
    \includegraphics[width=0.99\columnwidth]{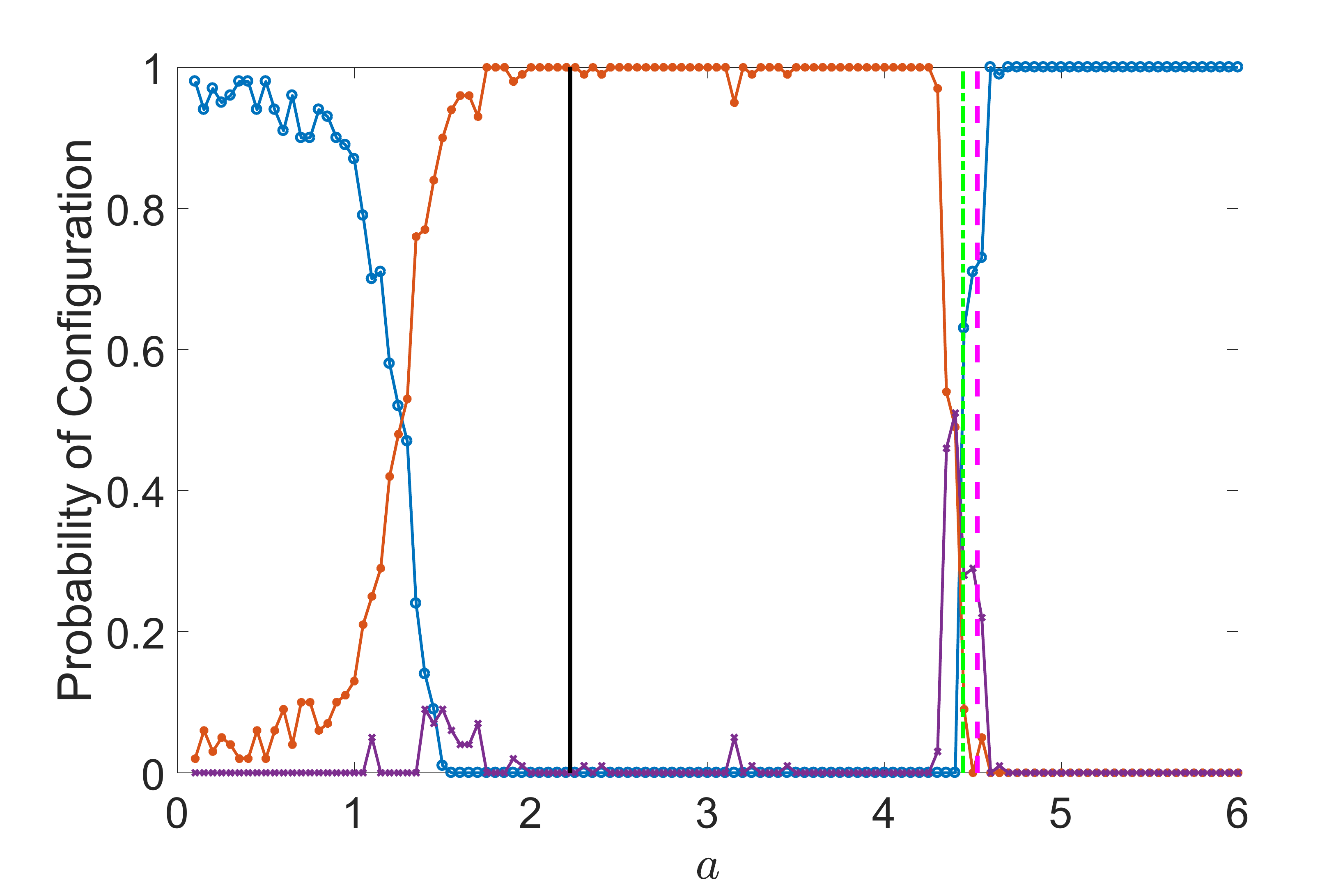}
    \caption{\textbf{Basins of attraction}. We plot the fraction of uniform random initial conditions that end up in bimodal (blue circles), trimodal (orange asterisks), or higher order multimodal (purple xs) states for the concrete system examined in the main text.  Here $N=100$, $K =-10000$ and oscillators' natural frequencies are drawn from the distribution $\mathcal{N}(0,100)$.  We performed 100 unique simulations for each value of $a$. Final states (presumed equilibria) were identified automatically via $k$-means clustering.  Thresholds given in the main text for stability of bimodality and the antiphase state are given by the solid black line and the dot-dashed green line, respectively. The threshold for the necessary condition for stability of the trimodal state is given by the vertical dashed magenta line. }
    \label{fig:BasinSize}
\end{figure}

\section{Basins of Attraction for Multimodal States}

We have conducted some preliminary numerical exploration of the sizes of basins of attraction for various equilibria for the example interaction function given in Eq.\;(11) of the main text. We simulated the system one hundred times with initial phases chosen independently at random from the uniform distribution over the circle, i.e. $\mathcal{U}(-\pi,\pi]$, and evaluated the fraction of the time that the system converged to each distinct equilibrium state.  Results are shown in Fig.~\ref{fig:BasinSize}, with $N = 100$, $K = -10000$, and oscillator natural frequencies drawn from the distribution $\mathcal{N}(0,100)$.

Fig.~\ref{fig:BasinSize} also shows the stability thresholds described in Eqns.~(12) (bimodal state), (A.13) (antiphase state), and trimodal state (A.12) of the main text, visualized by the solid black, and dot-dashed green, and magenta vertical lines respectively.  In order to classify the observed equilibria, we use a $k$-means algorithm on the unit circle, with the number of clusters, $k$, being decided by the gap statistic. We say that a equilibrium state is bimodal if $k=2$, trimodal if $k=3$, and so on. 

We note that the results are consistent with our analysis in that the probability of a configuration is always zero in ranges of $a$ where it is excluded.  Although, we have not analyzed equilibria with more than three modes, we observe that such modes are unlikely to be observed for most values of $a$, and thus have apparently small basins of attraction.

Given that this experiment was conducted with heterogeneous oscillators, this lends plausibility to the idea that the system will end up in a multimodal state for sufficiently large coupling. More formal analysis of the basin size of the bimodal and trimodal state will be left for future work.

\section{Critical Coupling Strength}

\begin{figure}[t!]
    \centering
    \includegraphics[width=\columnwidth]{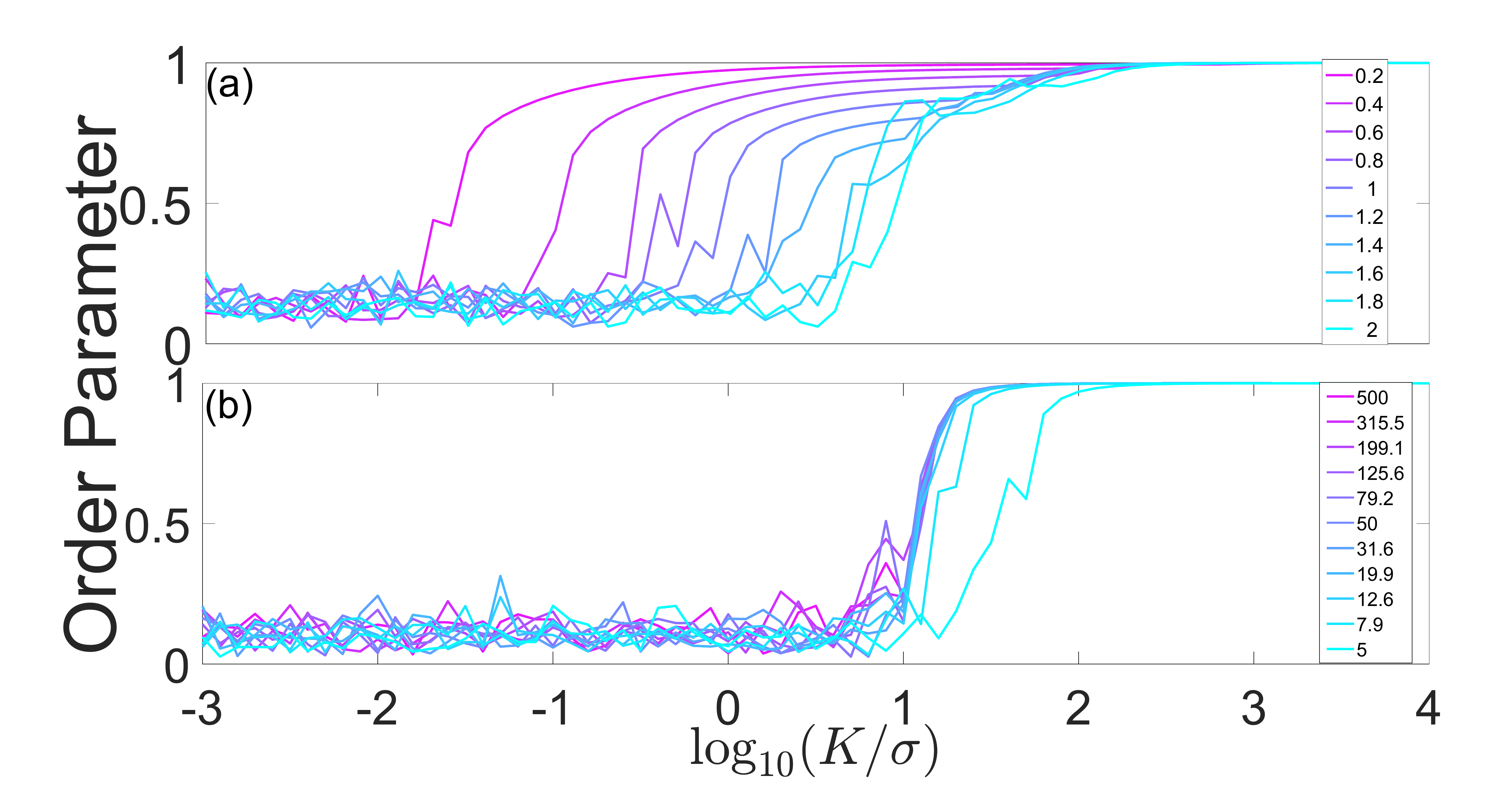}
    \caption{\textbf{Critical coupling strength}. We perform numerical experiments to demonstrate the existence of a critical coupling strength for our system and evaluate its dependence on parameter $a$ using the interaction function defined in Eq.~(11) of the main text. Here $N=100$, the natural frequency distribution is given by $\mathcal{N}(0,\sigma^2)$, and the initial phase distribution is $\rho(\theta) = 0.5\delta(\theta) +0.5 \delta(\theta-\psi_0)$, where $\psi_0$ is the predicted phase separation given by the stable fixed points of Eq.\;(11).  Here, each curve represents a different value of $a$ (values indicated in legend). As in the standard Kuramoto model, the critical coupling strength is dependant on the size of the standard deviation of the distribution, but unlike the standard Kuramoto model, it appears to also depend on $a$, which sets the shape of the interaction function. }
    \label{fig:CritKc}
\end{figure}

In the standard Kuramoto model with attractive coupling, there exists a critical coupling strength $K_c$ at which the system bifurcates from an incoherent state to the ordered state. To look for $K$ dependence in the system detailed in main text, we examine the simplest cases of $N=2$ and $N=3$, and also conduct several numerical experiments with results shown in Fig.~\ref{fig:CritKc}, though we leave more thorough exploration for future work. 

Figure \ref{fig:CritKc} shows how order varies as we increase coupling strength among nonidentical oscillators with the concrete interaction function used in the main text.  Here, we set $N = 100$ and draw the frequencies from the distribution $\mathcal{N}(0,\sigma^2)$. From here, we vary the quantity $K/\sigma$ so that $\log_{10}(K/\sigma)$ runs from -2 to 4. Each curves shown above represents the result of an experiment for a given value of $a$. Here, the order parameter is defined as follows:
\begin{equation}
 R = \max \left\{ \left\vert\sum_j \dfrac{e^{2i\theta_j}}{N} \right\vert,
 \left\vert\sum_j \dfrac{e^{3i\theta_j}}{N} \right\vert, 
 \left\vert \sum_j \dfrac{e^{\frac{2\pi}{a}i\theta_j}}{N}\right\vert  \right\}.
\end{equation}
Defining the order parameter in this fashion sets the value of the order parameter to be 1 whenever the final configuration is bimodal or an equally spaced trimodal solution.
Just as in the standard Kuramoto model, if the coupling strength $K$ is not sufficiently large in magnitude, the system goes to the incoherent state due to intrinsic oscillator heterogeneity. We observe that the critical coupling strength appears to be proportional to the standard deviation of the frequency distribution, similar to the result in the standard Kuramoto analysis, but we point out that the critical coupling strength $K_c$ also appears to have dependence on the value of $a$.  We believe that some insight into this dependence can be gained from examining the simple $N=2$ and $N=3$ cases, though more rigorous analysis is left for future work.

For $N=2$, the system reduces to

\begin{equation}
 \dot{\psi} = \Delta \omega - K f(\psi) \label{eq:psidot_hetero}
\end{equation}
where $\Delta \omega = \omega_2-\omega_1$. Setting $\dot{\psi} = 0$, we find that a fixed point $\psi_0$ must satisfy the equation:
\begin{equation}
\dfrac{\Delta \omega}{K} = f(\psi_0)\;. \label{eq:fp_hetero}
\end{equation}
Note, this fixed point does not always exist, but if the coupling function $f$ has zeros, a fixed point must arise as $|K| \rightarrow \infty$. 

Even without explicitly defining $\psi_0$, we can observe scaling dependencies for the critical coupling strength $K_c$, which is defined such that 
\begin{equation}
    f(\psi_{max}) = \dfrac{\Delta \omega}{K_c} \label{eq:coupstreng_cond}
\end{equation}
where $\psi_{max} \in (-\pi,\pi]$ is the value such that $f(\psi_{max}) = \max f(\psi)$ (the arg max). 
We observe that $K_c \propto \Delta \omega$, which is expected if $K_c \propto \sigma$ as in the standard Kuramoto model (since for two oscillators $\sigma \propto \Delta \omega$) and is observed in our numerical experiments even for $N \gg 2$. 

We also observe that $K_c$ scales with the maximum value of the interaction function $f$, which in our numerical experiments depends on the parameter $a$.  Similar dependence is also evident  if we consider the $N=3$ case.

For $N=3$, we take the natural frequencies (without loss of generality) to be $0, -\sigma/3, \sigma/3$ respectively. As before, we convert to difference coordinates $\psi_1 = \theta_2- \theta_1$ and $\psi_2 =\theta_3-\theta_2$, and arrive at two conditions for existence of equilibria:
\begin{align}
    \dfrac{\sigma}{K} = f(\psi_2-\psi_1) - f(\psi_2) -2 f(\psi_1) \\
    \dfrac{\sigma}{K} = f(\psi_2-\psi_1) + 2f(\psi_2) + f(\psi_1)\;, 
\end{align}
which simplify to 
\begin{align}
    \dfrac{\sigma}{K} &= f(\psi_2-\psi_1) + f(\psi_2) \\
    f(\psi_1) &= -f(\psi_2)\;.
\end{align}
Hence, a necessary condition $K$ must satisfy for the existence of equilibria is
\begin{equation}
     \dfrac{\sigma}{K} \leq 2 f(\psi_{max})\;.
\end{equation}
%
So, just as in the $N=2$ case, we see that the critical coupling strength $K_c$ is proportional to the oscillator heterogeneity $\sigma$ and inversely proportional to the maximum of the interaction function $f$.

We hypothesize that similar scaling laws hold for $N \gg 1$, and find that such a hypothesis is consistent with data from numerical experiments shown in Fig.~\ref{fig:CritKc}.

\end{document}